# Arc-to-pocket transition and quantitative understanding of transport properties in cuprate superconductors

W. Tabiś[1,2,3]*, P. Popčević[1,4]*, B. Klebel-Knobloch[1], I. Biało[1,2], C. M. N. Kumar[1,2], B. Vignolle[3,5], M. Greven[6], and N. Barišić[1,7]

**Despite immense efforts, the cuprate Fermi surface (FS) has been unambiguously determined in only two distinct, low-temperature regions of the phase diagram: a large hole-like FS at high doping, and a small electron-like pocket associated with charge-density-wave driven FS reconstruction at moderate doping. Moreover, there exists incomplete understanding of the reconstructed state, which is stabilized by high magnetic fields, and its connection with the normal state that consists of arc-like remnants of the large underlying FS. Part of the problem is that compound-specific idiosyncrasies, such as disorder effects and low structural symmetry, can obscure the fundamental properties of the quintessential $CuO_2$ planes. Here we present planar magnetotransport measurements for moderately-doped $HgBa_2CuO_{4+\delta}$ that enable a quantitative understanding of the phase transition between the normal and reconstructed states and of the charge transport in the latter, and that demonstrate that the quasiparticle scattering rate in both states is due to Umklapp scattering. Building on prior insights, we furthermore arrive at a comprehensive understanding of the evolution of the planar transport properties throughout the entire cuprate phase diagram.**

[1] Institute of Solid State Physics, TU Wien, 1040 Vienna, Austria

[2] AGH University of Science and Technology, Faculty of Physics and Applied Computer Science, 30-059 Krakow, Poland

[3] Laboratoire National des Champs Magnétiques Intenses (LNCMI-EMFL), (CNRS-INSA-UGA-UPS), Toulouse, 31400, France

[4] Institute of Physics, Bijenička cesta 46, HR-10000, Zagreb, Croatia

[5] Institut de Chimie de la Matière Condensée de Bordeaux, UMR 5026, Pessac, France

[6] School of Physics and Astronomy, University of Minnesota, Minneapolis, MN 55455, USA

[7] Department of Physics, Faculty of Science, University of Zagreb, Bijenička cesta 32, HR-10000, Zagreb, Croatia

*Equally contributing authors

*Introduction*

The lamellar cuprates exhibit a complex phase diagram as a result of strong electronic correlations, with an insulating state at zero doping, a Fermi-liquid (FL) state at high doping, and superconductivity as well as pseudogap (PG) and 'strange-metal' (SM) phenomena at intermediate doping (Fig. 1a). Depending on which of these regimes is taken as the starting point, seemingly mutually exclusive theoretical approaches have been developed,[1,2] including those evoking quantum criticality[3] and the holographic metals[4]. To a large extent, this longstanding debate has been fuelled by the unusual evolution of charge transport properties across the phase diagram.

At high hole-dopant concentrations $p$, the FS properties are well understood and described by the Landau FL formalism: for example, for $Tl_2Ba_2CuO_{6+\delta}$ (Tl2201), photoemission spectroscopy,[5] angle-dependent magnetoresistance[6] and quantum-oscillation (QO) experiments[7] reveal an approximately circular hole-like FS that occupies about 60% of the Brillouin zone, consistent with band-structure calculations.[8] The carrier density estimated from the Luttinger sum rule is $1 + p$ and agrees with low-temperature Hall-effect measurements.[9] Furthermore, for $c$–axis magnetic fields above the upper critical field, the Wiedemann-Franz law relating heat and charge transport is obeyed.[10] We refer to this state as the overdoped FL, or OD-FL.

With decreasing doping, electronic correlations give rise to the PG, a partial gap on the anti-nodal parts of the FS. This regime is marked by a number of unusual behaviours, including the linear temperature dependence of the planar resistivity, $\rho \propto T$. Nevertheless, below the characteristic temperature $T^{**}$ (PG-FL, Fig. 1a), the itinerant carriers associated with the arc-like FS remnants maintain FL character[11]: $\rho$ is quadratic in temperature and inversely proportional to $p$[12], whereas the Hall coefficient ($R_H$) is essentially temperature-independent and a good measure of $p$[11,13]; the magnetoresistivity obeys Kohler scaling, with a scattering rate that is quadratic in temperature, $1/\tau \propto T^2$;[14] the optical conductivity displays temperature-frequency scaling consistent with FL theory, with a carrier density that corresponds to $p$.[15,16] The existence of quasiparticles only on arc-like FS portions is highly unusual, and thus the PG-FL regime should be distinguished from the OD-FL. Additionally, this moderately underdoped part of the phase diagram exhibits charge-density-wave (CDW) correlations[17,18] which, in sufficiently high magnetic fields, induce a reconstruction of the arc segments into a small electron pocket[18,19,20,21]. The latter has been revealed *via* QOs and shown to occupy 2-3% of the original Brillouin zone (R-FL, Fig. 1c,d).[21,22,23,24]

These observations raise several important, interrelated questions: (1) Is it possible, as in the case of the simple OD-FL state, to reach a quantitative understanding of the charge transport in the reconstructed state? (2) What is the key relation between the charge transport in the PG-FL and R-FL states? (3) Can we comprehensively understand the evolution of the planar transport properties throughout the entire phase diagram? Our magnetotransport measurements for the model cuprate Hg1201 allow us to answer the first two questions. With FS parameters inferred from QO measurements, and assuming Umklapp as the dominant scattering process, FL theory predicts the absolute values and the temperature and magnetic field dependencies of the transport coefficients. We show that those values are in remarkable quantitative agreement with our experimental results. Consistent with prior work,[25] we find a resistivity peak concomitant with the sign change of the Hall coefficient, which may be attributed to a transition between the PG-FL and R-FL phases. For the R-FL state, we demonstrate Kohler scaling of the magnetoresistance and quantitative understanding of the Hall constant. We furthermore demonstrate that the same (Umklapp) scattering mechanism that characterises the OD-FL is at play in both underdoped phases, with scattering rates that naturally differ by an order of magnitude. Finally, by building on previous insights[11,26] and further utilizing predictions of FL theory, we arrive at an overall understanding of the charge transport across the cuprate phase diagram, and thereby answer the third question. We therefore resolve a major conundrum pertaining to these important quantum materials.

*Results*

Measurements of $\rho$ and $R_H$ were performed in pulsed magnetic fields up to 68 T in transverse geometry ($j \parallel ab$; $H \parallel c$; see Supplementary Section 1 for details). We present results for two underdoped Hg1201 crystals with similar zero-field critical temperatures, $T_c = 71$ K and 72 K, labelled UD71 and UD72, respectively. The samples were prepared and characterized following established procedures and exhibit hole concentrations of $p \approx 0.09$ and $p \approx 0.095$, respectively (see Supplementary Section 1). A detailed study of UD71 was performed; this sample was previously used for QO measurements,[22] and its doping level is close to where CDW correlations are most prominent[17,18] and, hence, where the FS reconstruction is expected to be most pronounced (Fig. 1a). Additional measurements of UD72 were performed to verify the robustness of the results.

Figure 2 shows the temperature and magnetic field dependence of the resistivity for UD71. The temperature dependence at fixed field (Fig. 2a) is obtained from isothermal magnetoresistivity measurements (Supplementary Section 1, Fig. S1). With increasing field, the superconducting phase is suppressed, and a peak emerges, with a maximum around 20 K. This feature becomes apparent at about 30 T and is more pronounced in higher fields. Such a temperature dependence is usually associated with a phase transition, *e.g.*, the formation of CDW order (Supplementary Section 2).

Figure 2b shows that, up to approximately 10 K, the high-field resistivity exhibits quadratic temperature dependence, indicative of FL behaviour. The concomitant quadratic field dependence, expected for a FL, is demonstrated in Fig. 2c,e. Back-extrapolation of $\rho(H)$ from above the upper critical field $H_{c2}$, determined as the field above which $\rho \propto H^2$, results in an estimate of the (hypothetical) zero-field resistivity $\rho(H=0)$ expected if the reconstructed state were to exist in zero field, in the absence of superconductivity. As seen from Fig 2d, $\rho(H=0) \propto T^2$ below 10 K, analogous to the temperature dependence of high-field resistivity (Fig. 2b).

Figure 3 compares the data for UD71 and UD72. For both samples, the resistivity maximum (Fig. 3a) corresponds rather well to the temperature ($T_{FSR}$) at which $R_H$ changes sign (Fig. 3b). For UD72, $T_{FSR} \approx 16$ K is about 4 K lower than for UD71, indicative of (expected) weaker CDW correlations at this higher doping level. $R_H$, which is approximately constant below $T^{**}$,[11,27] begins to decrease relatively steeply below ~40 K (Fig. 3b). Below ~5 K, $R_H$ reaches a constant value (Fig. 3b), as expected once the transition to the R-FL state is completed. Moreover, the field dependence of the Hall resistivity for UD71 (see Supplementary Section 1, Fig. S1c,d) clearly demonstrates that the applied fields are sufficiently large to entirely suppress superconductivity, since the high-field data linearly back-extrapolate to zero at 0 T. Notably, since $\rho \propto T^2$ in the same temperature range (Fig. 3c), the cotangent of the Hall angle (*i.e.*, the inverse Hall mobility), $\cot(\theta_H) = \rho/(HR_H)$, is also quadratic in temperature, as expected for a FL (Fig. 3d).

The data furthermore allow us to determine the temperature dependences of $H_{c2}$ and of the vortex-solid melting field, $H_{VS}$ (Fig. 4a,b). The latter is determined as the field above which the resistivity deviates from zero, *i.e.*, rises above the noise level. The estimation of $H_{c2}$ was possible only up to ~12 K because of the significant deviation from $\rho \propto H^2$ at higher temperatures due to the phase transition around 20 K and the appearance of superconducting correlations. Analysis for UD71, for which the more detailed data set was collected, reveals a non-monotonic temperature dependence of $H_{c2}$ that reflects the CDW-superconductivity competition (Fig. 4c,d; Supplementary Section 3).

*Discussion*

We now discuss the three pivotal, interrelated questions raised in the introduction: Can we reach quantitative understanding (1) of the charge transport in the R-FL state and (2) of the connection between PG-FL and R-FL states, and (3) is it possible to comprehensively understand the planar transport throughout the entire cuprate phase diagram?

**Quantitative understanding of the R-FL and OD-FL states.** Following the initial discovery of QOs in moderately-doped $YBa_2Cu_3O_{6+d}$ (Y123)[23] and $YBa_2Cu_4O_8$ (Y124)[24], two structurally complex cuprates ($CuO_2$ double-layers, Cu-O chains, low structural symmetry), the observation of QOs in simple-tetragonal Hg1201 demonstrated that this behaviour characteristic of a FL is a (universal) property of the quintessential $CuO_2$ planes[22]. Subsequent work for Hg1201 confirmed[21] the existence of a single electron pocket in the R-FL state[22] and revealed a quantitative relation between the pocket size, estimated arc-like FS, and the measured CDW wavevector[17,18]. We now demonstrate that other crucial aspects of the R-FL can be understood at a quantitative level and that this state obeys Kohler scaling.

FL theory gives a clear, quantitative prediction for the transport coefficients. The resistivity is quadratic in temperature: $\rho(T) = A_2 T^2 + \rho_{res}$, where $\rho_{res}$ is a residual contribution associated with crystal lattice imperfections. Given the lamellar structure of the cuprates, the *sheet resistance* should be scrutinized (*i.e.*, the resistance per $CuO_2$ plane, $\rho_\square = \rho/c^*$, where $c^*$ is the distance between neighbouring planes).[12] $A_2$ can be estimated from the expression for the self-energy of a FL;[28] for a two-dimensional (2D) single-band metal ($m^*$: effective mass, $k_F$: Fermi wave vector; subscript "t" indicates theoretical estimate):

$$A_{2\square,t} = \frac{9\pi^4 k_B^2}{e^2 \hbar^3} \frac{m^{*2}}{k_F^4} \tag{1}$$

A similar expression was developed earlier[29] and used to estimate $A_{2\square}$[30] (see Supplementary Section 4 for details).

In the OD-FL, where the carrier density is $1 + p$,[9] dissipation is understood to result from Umklapp scattering,[31] and QO measurements directly reveal key FS properties: for Tl2201 at $p \approx 0.30$, $m^* = 4.1 \pm 1.0$ $m_e$ and $k_F = 0.742 \pm 0.005$ Å$^{-1}$.[7] Assuming a 2D parabolic band (*i.e.*, a circular Fermi surface with radius $k_F$), consistent with experiment,[7] Eq. (1) yields $A_{2\square,t} = 0.025 \pm 0.010$ Ω/K$^2$, in good agreement with the experimental value $A_{2\square} = 0.022 \pm 0.007$ Ω/K$^2$.[32,33] Remarkably, the same arguments also yield quantitative understanding of the R-FL phase: in Hg1201, QO measurements[21,22] reveal a sole electron pocket with $m^* = 2.45 \pm 0.15$ $m_e$ and $k_F = 0.16 \pm 0.03$ Å$^{-1}$, and Eq. (1) yields $A_{2\square,t} = 4.2 \pm 0.6$ Ω/K$^2$. This value is in excellent agreement with our experimental result $A_{2\square} = 4.1 \pm 0.3$ Ω/K$^2$ (Fig. 2d).

Next, we demonstrate in Fig. 2f that the transverse magnetoresistance, $\rho(H)-\rho(H=0))/\rho(H=0)$, collapses to a single curve below 10 K for $H > H_{c2}$. This scaling demonstrates the validity of Kohler's rule, a key FL property, in the R-FL state. Figure 2f also shows the temperature dependence of the coefficient $a_\perp$, obtained from fits of the magnetoresistance to $a_\perp H^2$ (see Supplementary Section 5). Unsurprisingly, $a_\perp$ deviates from simple $1/T^4$ dependence and instead follows $a_\perp = (d + eT^2)^{-2}$, because $\rho_{res}$ is not negligible and the scattering rate ($\rho \propto 1/\tau$) contains this temperature-independent residual term, in accordance with Matthiessen's rule.[14] We find excellent agreement between the ratios $d/e = 250 \pm 20$ K$^2$ and $\rho_{res}/A_2 = 240 \pm 10$ K$^2$, the latter obtained from the resistivity data extrapolated to $H = 0$ T and $T = 0$ K (Fig. 2d), as expected if Kohler's rule is obeyed.

Furthermore, using the FS parameters extracted from QO experiments performed previously on UD71, and assuming a circular FS, results in the estimated Hall coefficient $R_H = -14.7 \pm 0.6$ mm$^3$/C.[22] This estimate agrees with our measured value $R_H = -14.5 \pm 0.5$ mm$^3$/C (Fig. 3b). This agreement implies that the Hall coefficient is a good measure of the carrier density in Hg1201, not only in the normal state (PG-FL, SM and OD-FL)[11], but also in the R-FL state. The negative sign of $R_H$ is consistent with the electron-like character of the reconstructed FS established for Hg1201 and Y123.[27,34]

**PG-FL versus R-FL.** Since $\rho \propto T^2$ in the R-FL phase (Figs. 2 and 3), $\cot(\theta_H) \propto T^2$ is expected in the temperature range where $|R_H|$ has levelled off (<5 K). This is demonstrated in Fig. 3d. However, the coefficient $\cot(\theta_H)/T^2 = 0.5 \pm 0.2$ K$^{-2}$ is about thirty times larger than the universal normal-state value $0.0175 \pm 0.0020$ K$^{-2}$,[11] indicative of dramatically different electronic scattering rates in the R-FL and (zero-field) normal states, consistent with a phase transition. This is also apparent from the temperature dependence of the resistivity. In the PG-FL state, $A_{2\square}/p = (9.0 \pm 2.7$ m$\Omega$/K$^2$)/p (Ref. 12) yields $A_{2\square} = 0.10 \pm 0.03$ $\Omega$/K$^2$ for $p = 0.09$, approximately forty times smaller than the R-FL state value $A_{2\square} = 4.1 \pm 0.3$ $\Omega$/K$^2$ (Fig. 2d). In contrast, the relatively small increase of $|R_H|$ in the R-FL phase suggests that approximately 1/3 of the doped holes are gapped as a result of the FS reconstruction. This corresponds to ~0.027 hole per planar CuO$_2$ unit at $p = 0.09$ and agrees well with estimates of the fractional electron charge (0.028) involved in the formation of the CDW order.[35,36]

High-field magnetoresistance measurements similar to the present work were previously conducted for Y124.[30] Although Y124 exhibits QOs[24] and a sign change of $R_H$ (at ~30 K),[34] no evidence for a phase transition was discerned from the temperature dependence of the resistivity. The structural complexity of Y124 noted above results in a complex FS that requires a multi-band description[30] and likely obscures the transition in resistivity measurements. Moreover, in structurally similar Y123, large $c$-axis magnetic fields induce additional, three-dimensional CDW correlations,[37,38] which are not likely the cause of the FS reconstruction, and might also be present in Y124 (see Supplementary Section 6). In contrast, Hg1201 is a tetragonal, single-CuO$_2$-layer cuprate without Cu-O chains and, unlike Y124 and Y123, exhibits a single QO frequency with high precision.[21] Hg1201 is therefore an ideal cuprate to document and understand the evolution from the normal (PG-FL) state to the R-FL state.

**Carrier-density evolution from OD-FL to PG-FL.** We have demonstrated that the FL formalism provides an accurate estimate of the resistivity coefficient $A_{2\square}$ in both the OD-FL and R-FL states. By following the doping dependence of $A_{2\square}$, we are now able to discuss the evolution of the FS across the phase diagram. Crucially, $\cot(\theta_H) = C_2 T^2 = \rho/(HR_H) \propto m^*/\tau$ (with $C_2 = 0.0175 \pm 0.0050$ K$^{-2}$) is nearly compound and doping independent, and thus invariant upon crossing from the OD-FL to the PG-FL (Fig. 1).[11,39] This universal behaviour eliminates the possibility of a conventional FS reconstruction between the two regimes, which would require changes in $k_F$ and $m^*$ and result in distinct values of $1/\tau$ and $C_2$. Consequently, the scattering rate (and $m^*$) is essentially compound and doping independent, and the Umklapp mechanism characteristic of the OD-FL remains the dominant scattering process at low doping levels (down to at least $p = 0.01$)[11]. A simple, logical consequence is that the changes in the resistivity result from changes in the carrier density (see Supplementary Section 7, Fig. S5), a possibility noted early on[40], but not seriously considered until recently[11,26,41,42]. This conclusion is consistent with photoemission spectroscopy results, which show the development of a partial gap (the PG breaks the large FS into Fermi arcs), but no sign of a reconstruction of the underlying FS that contains $1 + p$ states (Fig. 1b).[43] Instead, with decreasing doping, the density of states at the Fermi level is gradually depleted in the antinodal FS regions such that, in the PG-FL, only $p$ carriers of OD-FL character remain on the Fermi arcs.[11,12] This unusual $k$-space evolution has been attributed to an inhomogeneous Mott-like (de)localization of one hole per $CuO_2$ unit.[11,26]

We can therefore estimate $A_{2\square}$ for the PG-FL from the well-understood OD-FL value, accounting for the differences in carrier density[11,26] by simply multiplying $A_{2\square} = 0.022 \pm 0.007$ $\Omega/K^2$ for Tl2201 at $p = 0.3$ (Ref. 33) by $(1 + 0.3)/0.09$, which yields $A_{2\square} = 0.32 \pm 0.10$ $\Omega/K^2$ at $p = 0.09$. Since Tl2201 cannot be underdoped, we compare this result with the value $A_{2\square} = 0.10 \pm 0.03$ $\Omega/K^2$ at $p = 0.09$ determined from the experimentally-established universal relation.[12] Given the relatively large uncertainty in $A_{2\square}$ for Tl2201, this comparison is already quite satisfactory. However, we find even better agreement upon considering possible small FS differences: $C_2$ extracted from $\cot(\theta_H) = C_2 T^2$ is approximately 25% smaller for underdoped Hg1201 than for overdoped Tl2201, which indicates that $m^*$ for Hg1201 is approximately 25% smaller.[11] With this smaller value, Eq. (1) yields $A_{2\square,t} = 0.18 \pm 0.06$ $\Omega/K^2$ for Hg1201 ($p = 0.09$) which, within error, agrees with the experimental value $A_{2\square} = 0.10 \pm 0.03$ $\Omega/K^2$.[12]

Recent work revealed a peak-like resistivity feature similar to that in Fig. 2a and attributed this to a phase transition.[25] However, it should be noted that the temperature dependence of the carrier mobility, as suggested in the case of Y124,[30] could potentially result in various resistivity anomalies. The present study goes beyond these two works, as it establishes a quantitative understanding of the transport properties and connects various parts of the phase diagram. These connections also imply that, indeed, the observed transport-property changes are the result of the FS reconstruction. It is an intriguing question if, in the absence of both superconductivity and significant CDW correlations (and, hence, FS reconstruction), the arc metal persists to zero temperature. An interesting aspect of the data in Ref. 25 is that, for very underdoped Hg1201 ($p \approx 0.06$), the quadratic PG-FL temperature dependence seems to persist from $T^{**} \approx 230$ K down to the lowest measured temperature of a few Kelvin (Supplementary Section 8). Since static CDW correlations are absent at this doping level (Fig. 1), this indicates that the arc-like FS likely persists to zero temperature. Therefore, within increasing doping away from the Mott insulator, superconductivity fundamentally emerges from the arc metal at zero temperature, analogous to the emergence of superconductivity on cooling. Finally, we note that recent photoemission and QO measurements of five-layered $Ba_2Ca_4Cu_5O_{10}(F,O)_2$ at even lower doping indicate that, near the AF phase, the arcs are folded into hole-pockets by the strong AF correlations (see Supplementary Section 9),[44] analogous to the electron-pocket formation due to CDW order[17,18,19,21,22].

*Conclusions*

Our results for simple-tetragonal Hg1201 reveal profound connections among different parts of the cuprate phase diagram, which is enabled by the pivotal quantitative understanding of the R-FL state. Below about 7 K, in the R-FL state, the planar resistivity exhibits $T^2$ and $H^2$ dependences and obeys Kohler scaling, as expected for a FL. With $k_F$ and $m^*$ known from the prior QO work (performed on the same UD71 sample), we find that, just as for the OD-FL state, FL theory is in excellent agreement with the measured planar resistivity coefficient $A_{2\square}$. Moreover, the Hall coefficient (and hence the carrier density) is consistent with the QO result and independent of temperature (below 7 K), and the inverse Hall mobility follows $T^2$ dependence. Notably, the quasiparticle scattering rates in the PG-FL and R-FL phases differ by an order of magnitude and are consistent with the respective underlying FSs, as expected from the FL prediction $A_2 \propto 1/k_F^4$. From the Hall coefficient, we estimate that the FS reconstruction involves a loss of about 1/3 of the normal-state quasiparticles, fully consistent with prior NMR work.

The evolution of the transport coefficients across the cuprate phase diagram established here, and in particular the observation of significantly different scattering rates in the R-FL and PG-FL phases, implies that the previously-observed QOs correspond to a distinct low-temperature high-magnetic field phase. Thus, we provide crucial evidence that the resistivity peak and the sign-change of the Hall coefficient, at about 20 K, indeed correspond to a phase transition. The phase transition is associated with a FS reconstruction from arcs to an electron pocket due to bidirectional CDW order, in agreement with a range of earlier experimental results[17,18,19,20,21].

Remarkably, the planar charge-transport properties in all regimes of the hole-doped (and electron-doped[39,45]) cuprates can therefore be understood within the same conventional framework. This conclusion is of high significance, as it implies that there is no need to invoke exotic transport scattering mechanisms or quantum criticality. The transport properties directly stem from the Fermi-arc evolution with doping, temperature, and magnetic field, where arc states remain essentially unchanged, and from a scattering rate that is dominated by the Umklapp process. This means that the carrier density changes gradually as a function of temperate and doping and can be precisely determined from the resistivity[26] and, in relatively simple materials such as Hg1201, from the Hall effect.[11,46]

We emphasise that the cuprates are of course not simple Fermi-liquid metals, as these complex oxides are characterized by dual FL and Mott/PG physics, both in *k*-space and locally in real space, and by a high degree of inherent structural inhomogeneity. This complexity manifests itself in inhomogeneous real-space PGs associated with a gradual, percolative (de)localization of holes[11,26,42] and in the percolative emergence of superconductivity.[47,48]

*Author contributions*

MG, NB conceived the research. MG provided the samples. WT, PP, BKK, CMNK, NB characterized and prepared samples. WT, BV, NB performed the measurements. NB proposed the main lines of interpretation. WT, PP, IB conducted the analysis and calculations. WT, MG and NB wrote the paper with contributions from all authors.

*Acknowledgments*

We would like to thank O. S. Barišić, D. K. Sunko, D. Pelc and C. Proust for discussions and comments on the manuscript, and M. K. Chan for technical support with the measurements. The work at the TU Wien was supported by the European Research Council (ERC Consolidator Grant No. 725521), while the work at the University of Zagreb was supported by project CeNIKS co-financed by the Croatian Government and the European Union through the European Regional Development Fund-Competitiveness and Cohesion Operational Programme (Grant No. KK.01.1.1.02.0013). The work at AGH University of Science and Technology was supported by the Polish National Agency for Academic Exchange under 'Polish Returns 2019' Program, Grant No: PPN/PPO/2019/1/00014 and the subsidy of the Ministry of Science and Higher Education of Poland. The work at University of Minnesota was funded by the Department of Energy through the University of Minnesota Center for Quantum Materials, under Grant No. DE-SC-0016371.

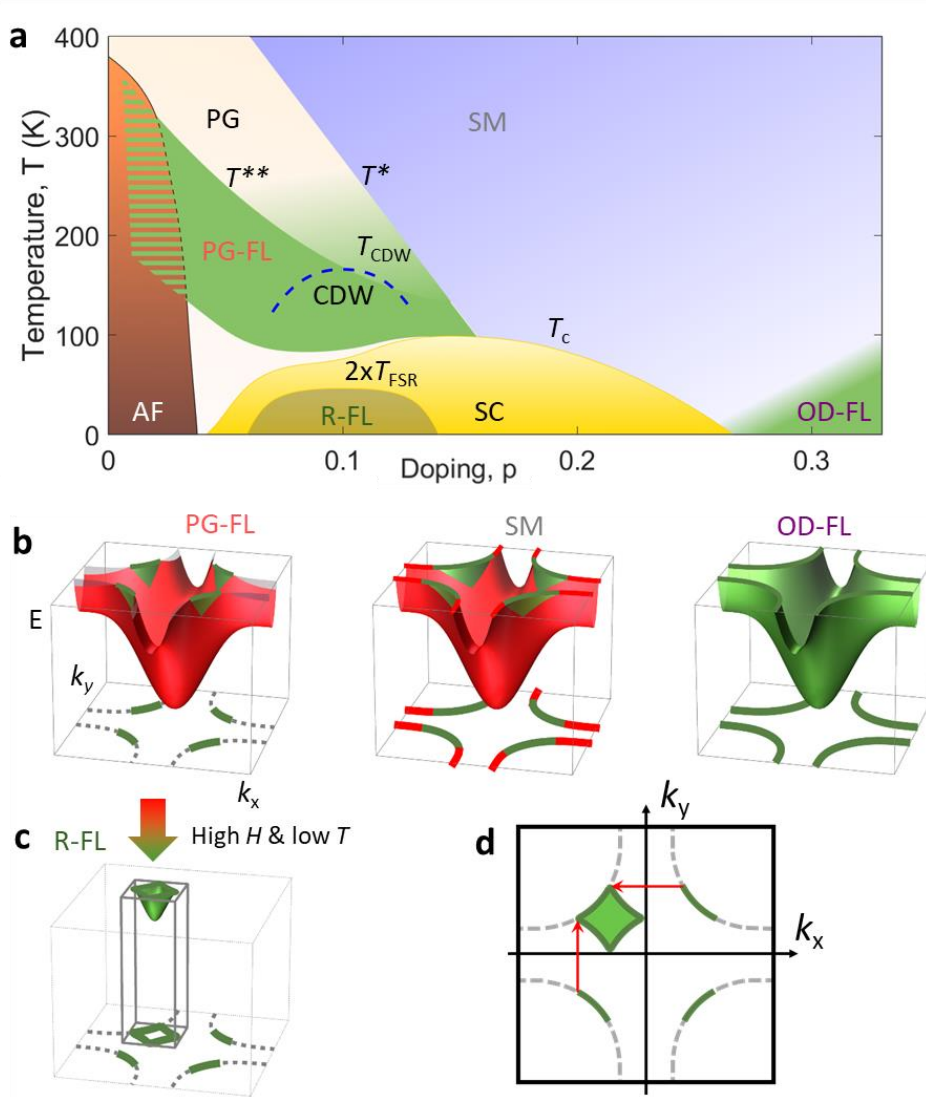

**Figure 1 | Schematic cuprate phase diagram with band dispersion. a,** Temperature-doping phase diagram, based on results for Hg1201 and Y123 (below $p \approx 0.20$)[11,18] and Tl2212 (above $p \approx 0.20$).[32,49] Green region at high doping: overdoped Fermi liquid (OD-FL) with carrier density $1 + p$.[9] A similar region (PG-FL) with carrier density $p$ exists below $T^{**}$.[12] SM: "strange-metal" regime, where the resistivity is approximately linear in temperature, but the scattering rate remains proportional to $T^2$.[11,12,26] SC: superconducting order. AF: antiferromagnetic order. $T_{CDW}$: onset of static short-range charge-density-wave (CDW) order.[18] $T_{FSR}$: corresponds to the maximum in resistivity (about 20 K for $p \approx 0.09$, not shown to scale), associated with the FS reconstruction, and manifests itself as a change in the sign of the Hall coefficient.[25] **b,** Schematic band dispersion and projected FS in the PG-FL, SM and OD-FL regimes inferred from transport and photoemission spectroscopy.[5,43,50,51] OD-FL shows FL character (green).[5,43] The PG forms with decreasing doping and temperature (red).[43,50] PG-FL: transport measurements, which probe an energy window $\sim k_B T$ around the Fermi level, reveal that the PG is fully developed along the antinodal FS portions (red)

and that the remaining nodal states (green) maintain (OD-)FL character.[11,50] **c,d** At low temperatures, once superconductivity is suppressed by a magnetic field, biaxial CDW order reconstructs the disconnected Fermi arcs into a small electron-like Fermi pocket (R-FL).[17,18,21] Grey lines: new reduced Brillouin zone. Using the underlying FS and measured CDW wavevector (red arrows), the predicted pocket size is in excellent agreement with QO measurements.[17,18,21]

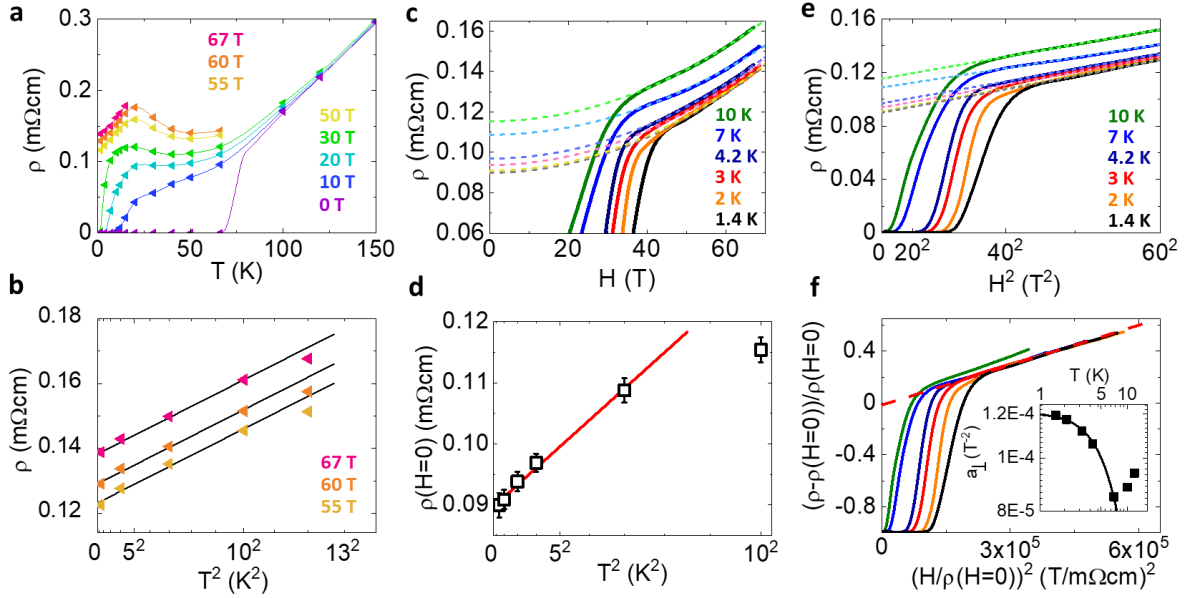

**Figure 2 | Magnetic field and temperature dependence of planar resistivity in Hg1201 UD71.**
**a,** $\rho(T)$ at indicated magnetic fields, obtained from isothermal resistivity data (Fig. S1). For $H >$ 20 T, a well-defined maximum develops at $T \approx 20$ K, indicative of a significant change in FS topology associated with the transition to the R-FL phase.[25] **b,** In the R-FL phase, $\rho(T)$ exhibits quadratic temperature dependence below ~10 K, indicative of FL behaviour. **c,** $\rho(H)$ at indicated temperatures. Dashed lines: quadratic fits to data in the R-FL state; extrapolation to $T = 0$ allows an estimate of $\rho(H=0)$. **d,** Temperature dependence of $\rho(H=0,T)$ with fit below 10 K to $\rho(H=0,T) = A_2T^2 + \rho_{res}$, where $A_2 = 3.9*10^{-9} \pm 0.3$ $\Omega$m/K$^2$ and $\rho_{res} = 897 \pm 5$ n$\Omega$m. With the lattice parameter $c = 9.6 \pm 0.1$ Å,[18] the sheet resistance coefficient is $A_{2\square} = 4.1 \pm 0.3$ $\Omega$/K$^2$. **e,** Low-temperature resistivity exhibits quadratic field dependence above the (temperature-dependent) $H_{c2}$. **f,** Demonstration of Kohler scaling in R-FL state (below ~ 7 K). Although $T^2$ and $H^2$ behaviour is observed up to 10 K (panels **b** and **e**), a small deviation from Kohler scaling is observed at this temperature: $|R_H|$ already has decreased by ~1/3 from its residual low-$T$ value (Fig. 3), indicative of a transition. Inset: Log-log plot of the transverse magnetoresistance coefficient $a_\perp$ (see Supplementary Section 4 for details). Black line: fit to $a_\perp = (d + eT^2)^{-2}$ below 10 K.

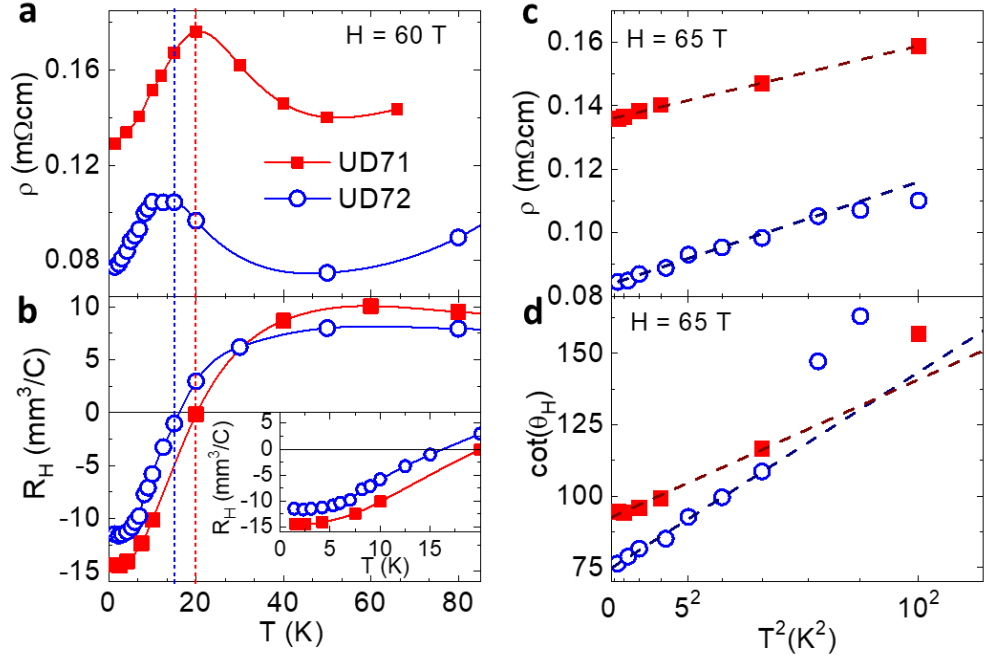

**Figure 3 | Temperature dependence of planar resistivity, Hall coefficient, and cotangent of Hall angle for Hg1201 UD71 and UD72. a**, Comparison of resistivity temperature dependence at 60 T. **b**, $R_H$ changes sign at $T_{FSR}$, approximately the same temperature at which $\rho$ (panel **a**) exhibits a maximum (vertical dashed lines). Below ∼5 K, $R_H$ is independent of temperature (inset), with a larger absolute value than in the PG-FL state. This indicates a depletion of approximately 1/3 of the quasiparticles as a consequence of the arc-to-pocket reconstruction. **c**, Comparison of the low-temperature resistivity at 65 T. The quadratic temperature dependence is observed in a smaller temperature range for UD72, consistent with the lower value of $T_{FSR}$, and indictive of somewhat weaker CDW correlations that UD71. **d**, For both samples, the cotangent of the Hall angle exhibits $T^2$ behaviour in the same temperature range in which Kohler's rule is obeyed (Fig. 2f). We obtain $C_2 = \cot(\theta_H)/T^2 = 0.49 \pm 0.05$ K$^{-2}$ and $0.68 \pm 0.03$ for UD71 and UD72, respectively, more than an order of magnitude larger than the universal value $C_2 = 0.0175 \pm 0.0020$ K$^{-2}$ in the PG-FL, SM and OD-FL regimes.[11] All lines are guides to the eye.

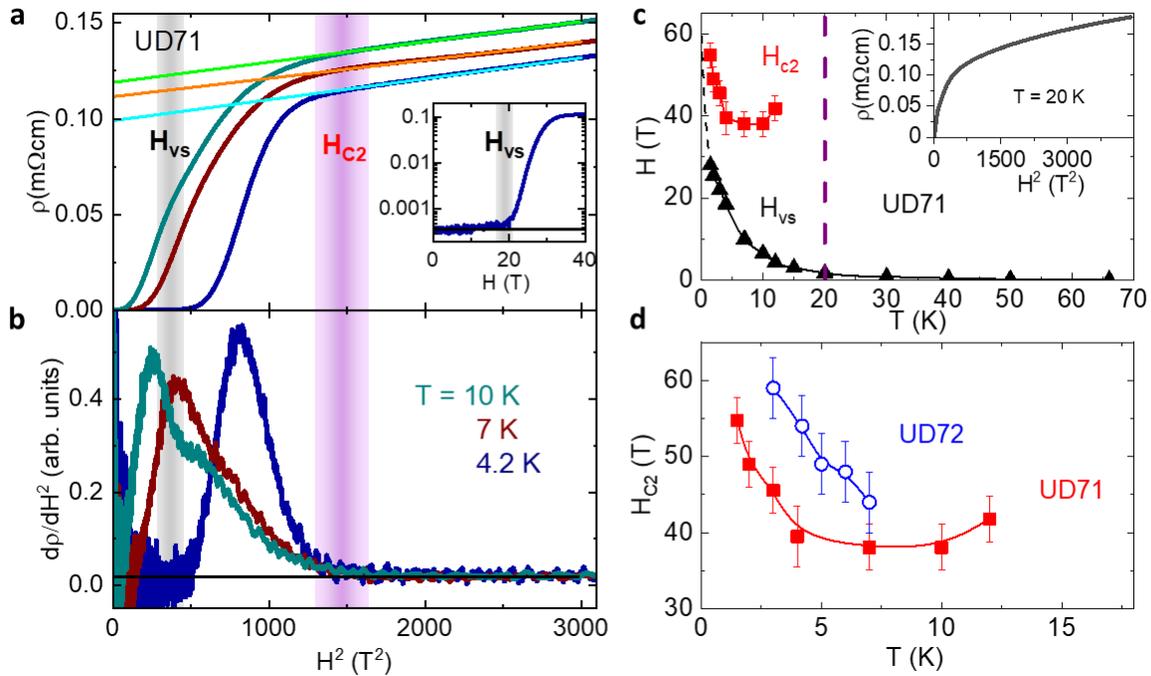

**Figure 4 | Critical fields for Hg1201. a**, Magnetoresistivity as a function of $H^2$ for UD71 at 10 K, 7 K, and 4.2 K: $H_{vs}$ is the irreversibility field, defined as the field above which the sample exhibits measurable nonzero resistivity (grey vertical band for the 4.2 K data); the estimation of $H_{vs}$ is shown in the inset, where $\rho$ is plotted on a logarithmic scale. $H_{c2}$ is the upper critical field, above which the magnetoresistivity exhibits normal-state $H^2$ dependence. Between 4.2 and 10 K, $H_{c2}$ is approximately temperature-independent (vertical purple band). **b**, Derivative $d\rho/dH^2$ for the data in **a**. $H_{c2}$ marks the field above which this derivative approaches zero. **c**, Temperature dependence of $H_{c2}$ and $H_{vs}$ for UD71. At low temperatures, both $H_{vs}$ and $H_{c2}$ decrease sharply, but $H_{c2}$ levels off above approximately 4 K, and then becomes undefined above 15 K, because $\rho$ departs from the $H^2$ dependence (inset) in the vicinity of the phase transition. The temperature (20 K) at which $R_H$ changes sign for UD71 is marked by the purple dashed line. **d**, Comparison of $H_{c2}$ for both samples. UD72, the sample with the larger $H_{c2}$, exhibits the transition to the R-FL state at a lower temperature, consistent with the lower $T_{FSR}$ and somewhat weaker CDW correlations at this doping level.[18] Solid lines in **c** (main panel) and **d**: guides to the eye. Dashed black line in **c**: extrapolation of $H_{vs}$ to 0 K.

# SUPPLEMENTARY INFORMATION

W. Tabiś, P. Popčević, B. Klebel-Knobloch, I. Biało, C. M. N. Kumar, B. Vignolle, M. Greven, and N. Barišić

**Content:**

**S1. Sample preparation, experimental setup, and additional data**

**S2. Resistivity in canonical CDW systems**

**S3. Competition between CDW order and superconductivity**

**S4. Theoretical estimate of the electrical resistivity coefficient $A_{2\square,t}$**

**S5. Zero-field resistivity estimation and Kohler's rule**

**S6. Magnetic field induced three-dimensional CDW order**

**S7. Charge localization and the phase diagram**

**S8. Resistivity and Hall coefficient in very underdoped Hg1201**

**S9. Arc-to-hole-pocket FS transformation**

**(S1) Sample preparation, experimental setup, and additional data**

Hg1201 single crystals were grown by a two-step flux method[52] and subsequently annealed[53] to achieve the desired doping level. Two crystals with mid-point $T_c$ values of 71 K (UD71) and 72 K (UD72) and estimated[54] hole concentrations of $p \approx 0.090$ and $\approx 0.095$ were selected for the present study. The 10-90% width of the superconducting transition, as determined from magnetic susceptibility measurements, was in the 2-3 K range. The samples were then cleaved to obtain fresh, clean *ac*-surfaces onto which gold pads were evaporated. Wires (20 µm diameter) were attached with silver paste, and subsequently cured for several minutes under the same conditions as the prior long-term anneal, and then quenched to room temperature. The resistance of the contacts did not exceed 1 Ohm. The UD71 sample was used in a previous experiment that established Schubnikov-de Haas oscillations in the R-FL state.[22] The measurements were performed at the Laboratoire National des Champs Magnétiques Intenses (LNCMI) in Toulouse, France. Data were collected during the entire duration of the magnetic field pulse, and the resistivity values obtained on the rise and fall of the pulse were in good agreement, which excludes the possibility of heating due to Eddy currents. Transverse (Hall) resistance data were collected simultaneously with the longitudinal resistance on a pair of contacts perpendicular to the excitation current. The in-plane magnetoresistivity, Hall resistance, and Hall coefficient are presented in Fig. S1 and Fig. S2 for UD71 and UD72, respectively.

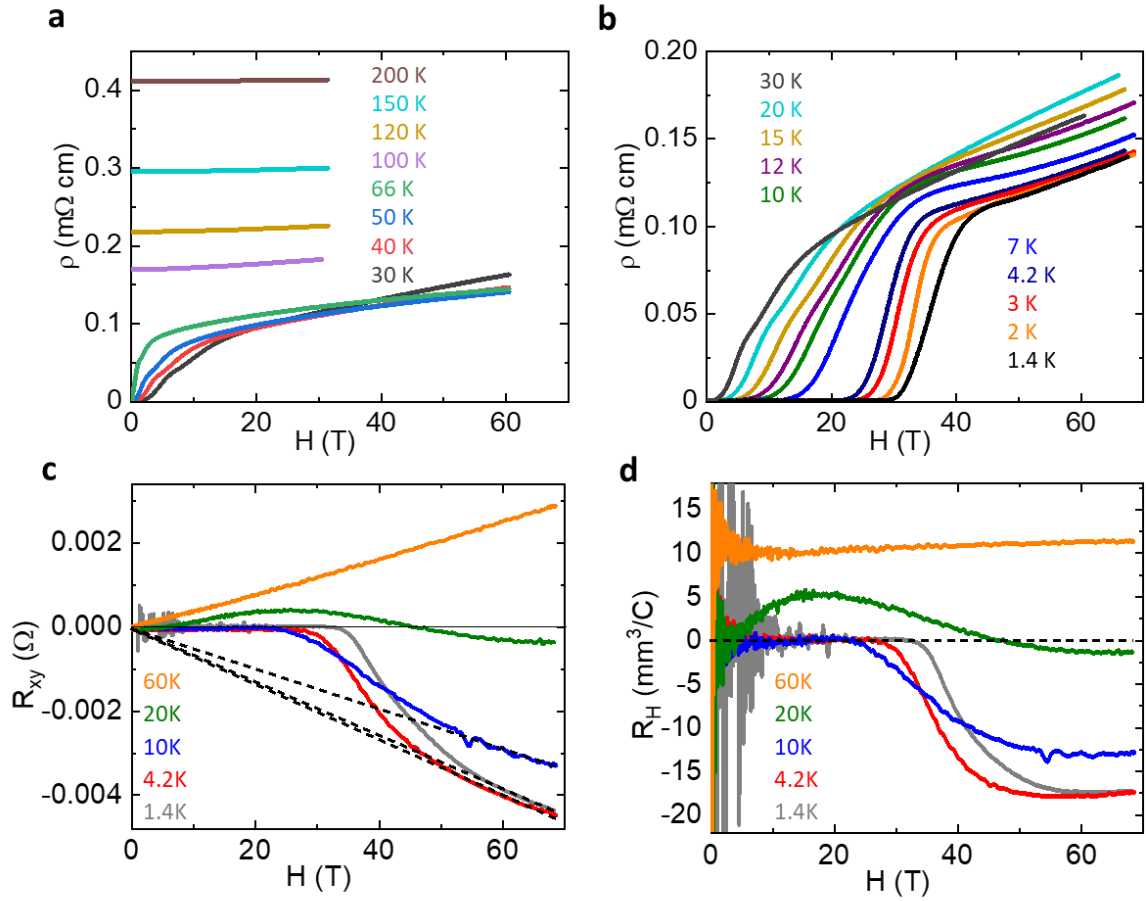

**Figure S1 | In-plane magnetoresistivity, Hall resistance, and Hall coefficient in Hg1201 UD71.** Complete isothermal magnetoresistivity data in the **a,** 30 to 200 K and **b,** 1.4 to 30 K ranges. These data were used to obtain the temperature dependence of the magnetoresistivity in Fig 2. **c,** Field dependence of transverse resistance $R_{xy}$ at constant temperature. Black dashed lines represent linear back-extrapolation from the high-field R-FL state data above $H_{c2}$. The Hall resistance vanishes in the limit $H \rightarrow 0$. **d,** Hall coefficient as a function of magnetic field, calculated from the isotherms of the transverse resistance in **c**, according to $R_H = w \cdot \frac{R_{xy}}{H}$, where $w$ is the sample width.

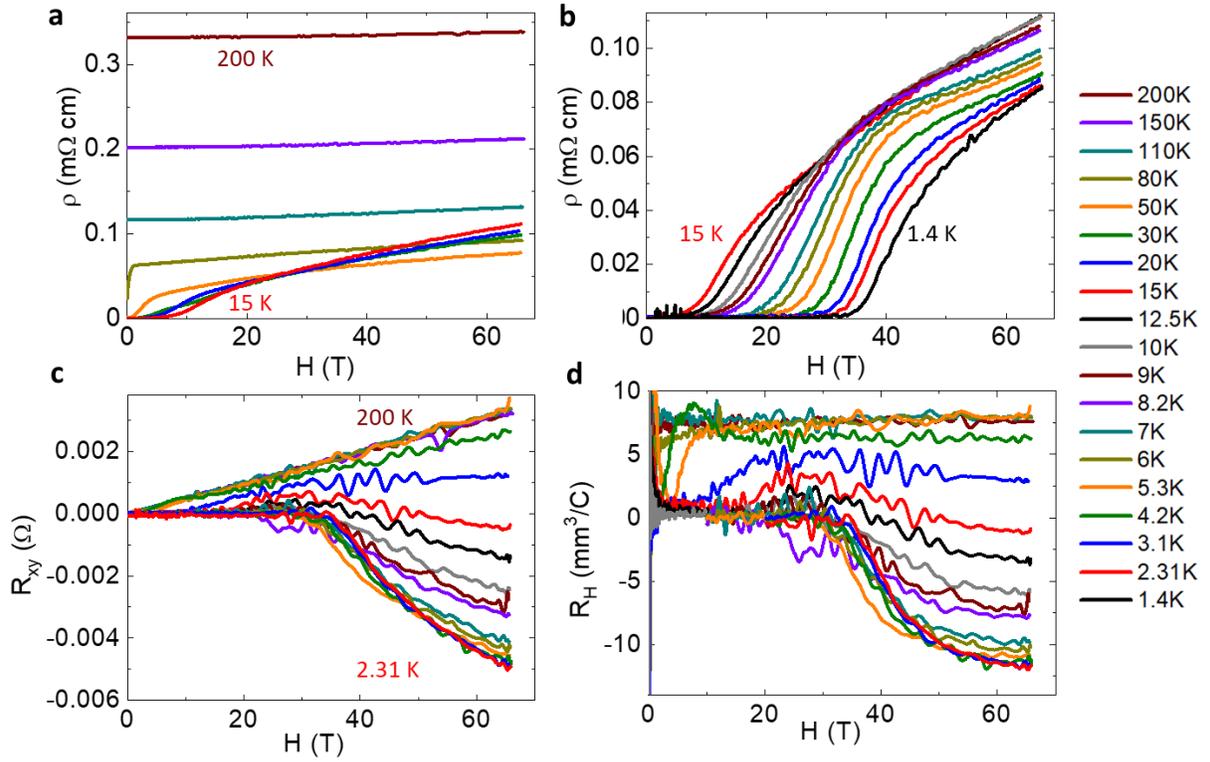

**Figure S2 | In-plane magnetoresistivity, Hall resistance, and Hall coefficient in Hg1201 UD72.** Complete isothermal magnetoresistivity data in the **a,** 15 to 200 K and **b,** 1.4 to 15 K ranges. **c,** Field dependence of transverse resistance $R_{xy}$ in the 2.31 to 200 K range. **d,** Hall coefficient as a function of magnetic field, calculated from the isotherms of the transverse resistance in **c**.

### (S2) Resistivity in canonical CDW systems

An excess of electrical resistivity is often a signature of an instability associated with a broken-symmetry state. Numerous CDW systems, *e.g.*, the quasi-one-dimensional conductors $A_xNb_3Te_4$ (A = Ag, Tl, In; Refs. 55, 56), quasi-two-dimensional purple bronzes $A_{0.9}Mo_6O_{17}$ (A=Na, K; Ref. 57), and the two-dimensional metal dichalcogenides $Pd_xTaSe_2$ (Ref. 58) and $TiSe_{2-x}S_x$ (Ref. 59) typically exhibit a resistivity increase already well above the transition temperature. For example, in $Nb_3Te_4$, CDW order sets in at 80 K,[60] as discerned via electron diffraction, although a resistivity increase can be easily observed already above 100 K (Fig. S3a, curve a). The transition can be shifted to higher and lower temperatures upon intercalation of the metal cations Tl or In (Fig. S3a,b). In particular, the CDW formation in $In_{0.75}Nb_3Te_4$ was observed by electron diffraction at ~140 K,[61] which corresponds to the temperature of the resistivity maximum (Fig. S3b).

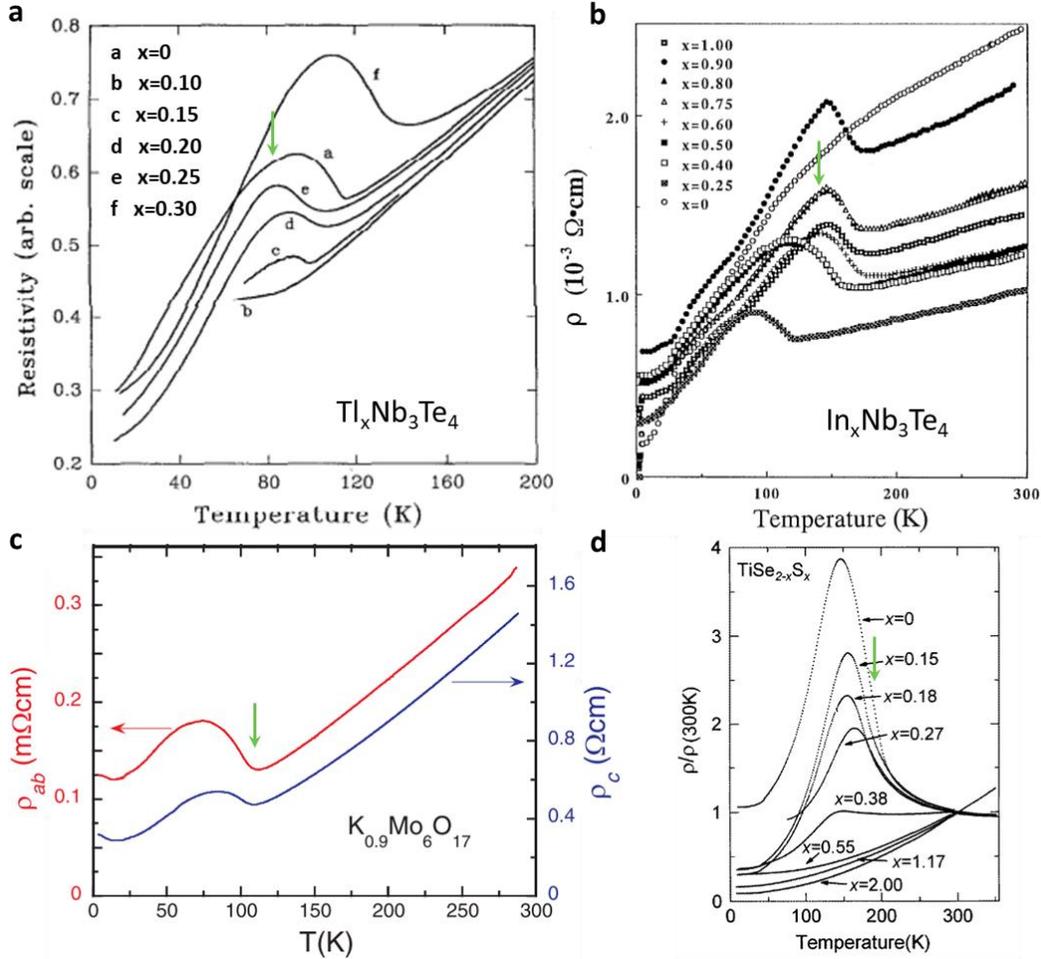

**Figure S3 | Temperature dependence of resistivity in various CDW compounds in the vicinity of the phase transition. a,** Resistivity of $Tl_xNb_3Te_4$ normalized to unity at 300 K (adapted from Ref. 55). In the host compound ($x = 0$), CDW order occurs at $T_{CDW} = 80$ K.[60] **b,** Resistivity of $In_xNb_3Te_4$ ($0 \leq x \leq 1.0$) observed in cooling measurements.[56] **c,** In-plane (left axis) and interplane (right axis) resistivity of $K_{0.9}Mo_6O_{17}$.[57] Both curves display an increase associated with the transition, which develops below $T_{CDW} = 115$ K.[62] **d,** Resistivity normalized at 300 K for $TiSe_{2-x}S_x$ (Ref. 59). Green arrows: $T_{CDW}$ for those sample compositions for which the transition was discerned by other probes. We note that, when analysing resistivity in CDW systems, it typically is not clear which criterion (*e.g.*, peak position, steepest slope) corresponds to the transition temperature determined from order parameter measurements.

Qualitatively, the resistivity anomaly that we observe for Hg1201 when superconductivity is suppressed by high magnetic fields (Figs. 2a, 2b, and Ref. 25) resembles that of the low-dimensional compounds, in particular that of the layered purple bronze $K_{0.9}Mo_6O_{17}$ (Fig. S3c; $T_{CDW} = 115$ K)[62], the hallmark of "hidden nesting" associated with a strongly anisotropic FS. However, lattice distortion does not develop in $K_{0.9}Mo_6O_{17}$, despite the commensurate CDW wave vector ($\mathbf{q}_{CDW}$), and none of the Fermi sheets are individually nested by $\mathbf{q}_{CDW}$.[62] A number of experiments (photoemission spectroscopy, Raman spectroscopy, low-energy electron diffraction, and x-ray scattering) favour a purely electronically-driven CDW instability in this compound.[62]

Another similar example is the lamellar material TiSe$_{2-x}$S$_x$ ($x = 0$), where the CDW phase transition occurs at $T_{CDW} \approx 200$ K,[63] marked by a strong resistivity peak at somewhat lower temperature (Fig. S3d). A large precursor regime is easily observable up to the room temperature. The flattening of the peak in $\rho(T)$ with increasing $x$ is attributed to a suppression of the CDW gap.[59]

**(S3) Competition between CDW order and superconductivity**

Examination of the $R_H$ isotherms for UD71 (Fig. S1c,d) provides additional information about the interplay between the PG-FL, SC, and R-FL phases. At 20 K, with increasing magnetic field, $R_H$ first reaches a local maximum at ~20 T as superconductivity is gradually suppressed, before changing sign above ~45 T. This behaviour is consistent with the prior observation for a Hg1201 sample with somewhat lower doping ($p \approx 0.075$, $T_c \approx 65$ K), where, at 10 K, $R_H$ was seen to increase with field, reach maximum at ~35 T, and then decrease at higher fields.[27] This nonmonotonic evolution of $R_H$ indicates that the transition to the R-FL phase requires a magnetic field, which promotes CDW order at the expense of superconducting order. Accordingly, 35 T is insufficient to reach the R-FL state. Consequently, the specific-heat measurements in ref. 64, aimed to measure the density of states, correspond to the complicated transient regime, and not to the pristine reconstructed phase.

Thermal conductivity measurements for Y123 indicate that, in the doping range relevant to our study, $H_{c2}$ decreases with increasing temperature in a rather conventional manner.[65] However, as shown in Fig. 4c, we find for Hg1201 UD71 that $H_{c2}$ (and $H_{vs}$) first decreases sharply with temperature and extrapolates to zero well below $T_c$. Thus, $H_{c2}$ does not exhibit conventional temperature dependence, which we attribute to the CDW-superconductivity competition. Instead, $H_{c2}$ levels off between 5 and 10 K and might even exhibit a local minimum; we are unable to follow the evolution above 15 K, since $\rho$ departs from $H^2$ dependence in the precursor regime.

As seen from Fig. 4d, $H_{c2}$ is larger for UD72 than for UD71 as a result of weaker CDW correlations in the former,[18] consistent with the fact that the signatures of FS reconstruction in this sample appear at a lower temperature (Fig. 3a,b). Finally, we note that the two samples display small differences in the slopes of the quadratic temperature dependences (Fig. 3c,d): $A_2$ ($H = 65$ T) $= 2.4*10^{-9} \pm 0.2$ $\Omega$m/K$^2$ and $3.4*10^{-9} \pm 0.2$ $\Omega$m/K$^2$ and $C_2 = \cot(\theta_H)/T^2 = 0.49 \pm 0.05$ K$^{-2}$ and $0.68 \pm 0.03$ for UD71 and UD72, respectively. These differences can be attributed to an increase of $m^*$ in the R-FL state toward optimal doping, as for example, suggested in the case of Y123.[66]

## (S4) Theoretical estimate of the electrical resistivity coefficient $A_{2\square,t}$

Metals in the FL regime exhibit a quadratic temperature dependence of the electrical resistivity, $\rho(T) = \rho_{res} + A_2 T^2$, and a linear temperature dependence of the electronic specific heat, $C_{el}(T) = \gamma T$. It was shown that the Kadowaki–Woods ratio (KWR), defined as $A_2/\gamma^2$, is universal for a variety of transition metals[67] and for heavy-fermion compounds,[68] yet with values that differ by nearly two orders of magnitude for those two material classes; for some other materials, this ratio was found to differ even more.[28] Subsequent considerations demonstrated that the KWR is independent of strong correlation effects, and only depends on non-renormalized material-specific parameters, such as the electron density, the density of states, and the Fermi velocity of the non-interacting system.[28] The dramatically different KWR for different materials classes was thus successfully clarified on the basis of material-specific parameters.

Assuming a momentum-independent phenomenological form for the self-energy of a local, momentum-independent Fermi liquid (introduced earlier in Ref. 69), the following fundamental ratio was proposed, which includes the effects of carrier density and spatial dimensionality:[28]

$$\frac{A_2 \, f_{dx}(n)}{\gamma^2} = \frac{81}{4\pi \hbar \, k_B^2 \, e^2} \tag{S1}$$

or, equivalently,

$$A_2 = \frac{81}{4\pi \hbar \, k_B^2 \, e^2} \frac{\gamma^2}{f_{dx}(n)} \tag{S2}$$

where $f_{dx}(n) \equiv nD_0^2 \langle v_{0x}^2 \rangle \xi$ may be written in terms of the dimensionality $d$ of the system, the electron density and, in layered systems, the interlayer spacing or the interlayer hopping integral. Here, $n$ is the conduction electron density, and $D_0$ is the bare density of states (DOS) at the Fermi level. $v_{0x} = \frac{\hbar^{-1}\partial \varepsilon_0(\mathbf{k})}{\partial k_x}$ is the bare velocity in the $x$ direction, while $\langle v_{0x} \rangle \langle \cdots \rangle$ denotes averaging over the Fermi surface. $\xi \approx 1$ is a number defined as: $1 < 2\xi \equiv 1 + \int_1^\infty y^{-2} F(y) dy \leq 1 + \int_1^\infty y^{-2} dy = 2$ as $F(y) \leq 1$ for $y \geq 1$, where $F(y)$ is a function related to the Kramers-Kronig relation applied to the real part of the self-energy $-\Sigma'(\omega, T)$.[28]

Furthermore, the total electronic heat capacity is given as:

$$\gamma = \gamma_0 \left(1 - \frac{\partial \Sigma'}{\partial \omega}\right) \tag{S3}$$

where $\gamma_0 = \frac{\pi^2 k_B^2 D_0}{3}$ is unrenormalized heat capacity coefficient, and $\left(1 - \frac{\partial \Sigma'(\omega,0)}{\partial \omega}\right)|_{\omega=0} = Z^{-1}$ is a renormalization factor due to strong interactions, $D_0$ represents the bare density of states, while $D^* = D_0/Z$ is the renormalized one.

For two-dimensional (2D) systems with only one band crossing the Fermi level, based on the dimensional analysis, one may introduce $m_0$ as a parameter that characterizes the bare density of states, $D_0 = m_0/\pi c \hbar^2$. Now, $m_0$ is the only parameter that can be renormalized by interactions. Thus, in the 2D single-band case Eq. (S3), can be rewritten as: $\gamma = \frac{\gamma_0}{Z} = \frac{\pi^2 k_B^2 D_0}{3Z} = \frac{\pi^2 k_B^2 m_0}{3 c \hbar^2 Z} = \frac{\pi^2 k_B^2 m^*}{3 c \hbar^2}$.

As further shown in Ref. 28, $f_{dx}(n)$ in Eq. (S2) takes the form $f_{2\parallel} = n^2/\pi c \hbar^2$, for the 2D case with charge transport in the direction parallel to the layers, and $k_F = \sqrt{2\pi c n} \Rightarrow n = k_F^2/2\pi c$, where $k_F$ is the Fermi wave vector.

Now, inserting the expressions $\gamma = \frac{\pi^2 k_B^2 m^*}{3 c \hbar^2}$ and $f_{2\parallel} = \frac{n^2}{\pi c \hbar^2}$ with $n = \frac{k_F^2}{2\pi c}$ into Eq. (S2) yields:

$$A_2 = \frac{9\pi^4 k_B^2 c}{e^2 \hbar^3} \frac{m^{*2}}{k_F^4} \tag{S4}$$

In the case of a single-band, 2D metal, and in terms of sheet resistance $\rho_\square = \frac{\rho}{c}$, where $c$ is the interlayer spacing, one obtains Eq. (1) in the main text:

$$A_{2\square,t} = \frac{9\pi^4 k_B^2}{e^2 \hbar^3} \frac{m^{*2}}{k_F^4} \tag{S5}$$

In 2005, Hussey[29] adopted a slightly different approach, starting with a phenomenological effective Fermi-liquid transport scattering rate, but with a similar ultimate understanding of the KWR. According to this approach, for a 2D metal, the sheet resistance coefficient (denoted as $A_{2\square,H}$ here) is:

$$A_{2\square,H} = \frac{8\pi^3 k_B^2 a}{e^2 \hbar^3} \frac{m^{*2}}{k_F^3} \tag{S6}$$

where $a$ is the planar lattice constant. Thus, the main difference between Eq. (S5) (which is the same as Eq. 1) and Eq. (S6) is that one factor of $k_F$ in the former is replaced by its limiting value $\frac{\pi}{a}$.

The theoretical predictions from Eqs. 1 (S5) and S6 are in good agreement: First, for overdoped Tl2201 (with $m^* = 4.1 \pm 1.0\ m_e$ and $k_F = 7.42 \pm 0.05$ nm$^{-1}$), we obtain $A_{2\square,H} = 0.021 \pm 0.01\ \Omega/K^2$ compared to $A_{2\square,t} = 0.025 \pm 0.01$ Ω/K². In the R-FL state of underdoped Hg1201 ($m^* = 2.45 \pm 0.15\, m_e$ and $k_F = 0.16 \pm 0.03$ Å⁻¹), Eq. (S6) gives $A_{2\square,H} = 0.75 \pm 0.1$ Ω/K². However, it is important to take into account the fact that the FS is reconstructed by the CDW order, and that the reconstruction leads to a reduced Brillouin zone. For Hg1201 ($p = 0.09$), the CDW wave vector is $q_{CDW} = 0.28 \pm 0.005$ r.l.u.,[17] which yields the reconstructed lattice constant $(3.57 \pm 0.06)a$, and hence $A_{2\square,H} = 2.7 \pm 0.4$ Ω/K². This value is consistent with, although about 25% lower than the values estimated from Eq. (1), $A_{2\square,t} = 4.2 \pm 0.6$ Ω/K², and from our data, $A_{2\square} = 4.1 \pm 0.3$ Ω/K². The values of the estimated coefficients are summarised in Tab. S1

| Parameter | OD-FL (Tl2201) | R-FL (Hg1201) |
|---|---|---|
| $m^*$ (QO) | $4.1 \pm 1.0\, m_e$ (Ref. 7) | $2.45 \pm 0.15\, m_e$ (Ref. 22) |
| $k_F$ (QO) | $7.42 \pm 0.05$ nm⁻¹ (Ref. 7) | $0.016 \pm 0.003$ nm⁻¹ (Ref. 22) |
| $A_{2\square,t}$ (calculation) | $0.02 \pm 0.01$ Ω/K² | $4.2 \pm 0.6$ Ω/K² |
| $A_{2\square}$ (experiment) | $0.022 \pm 0.007$ Ω/K² (Ref. 33) | $4.1 \pm 0.3$ Ω/K² |
| $R_H$(QO, circ. FS) | $1.65 \pm 0.015$ mm³/C | $-14.7 \pm 0.6$ mm³/C (Ref. 22) |
| $R_H$(Hall effect) | $1.0 \pm 0.2$ mm³/C (Ref. 46) | $-14.5 \pm 0.5$ mm³/C |
| $C_2$ | $0.017 \pm 0.002$ K⁻² (Ref. 11) | $0.49 \pm 0.03$ K⁻² |

**Table S1 | Parameters for the OD-FL state of Tl2201 ($p \approx 0.3$) and R-FL state of Hg1201 ($p = 0.09$).** The effective mass, $m^*$, and Fermi vector, $k_F$, are extracted from QO measurements. The values of $A_{2\square,t}$ are calculated from Eq. 1 and compared with the experimental values $A_{2\square}$. $R_H$ (QO, circ. FS) is estimated from QO measurements, assuming a circular FS. These values are compared with the results obtained from Hall-effect measurements, $R_H$ (Hall effect). $C_2$ is defined as the slope of the quadratic term in the expression $\cot(\theta_H) = C_0 + C_2 T^2$.

### (S5) Zero-field resistivity estimation and Kohler's rule

The zero-field resistivity, $\rho(H=0)$, is proportional to the scattering rate, $1/\tau$. If the carrier density and effective mass are constant, these two quantities are expected to exhibit the same temperature dependence in a conventional single-band metal. Furthermore, in a sufficiently weak applied magnetic field, the isothermal deviation of the resistivity from its zero-field value, $\delta\rho = \rho(H) - \rho(H=0)$, will follow a functional relation known as Kohler's rule: $\delta\rho/\rho(H=0) = F(H/\rho(H=0))$, where F is a scaling function. This follows from the fact that the field enters Boltzmann's equation in the combination $H\tau$. In most simple metals, the magnetoresistivity exhibits $H^2$ behaviour in the weak-field regime, so that Kohler's rule takes on the form:

$$\frac{\delta\rho}{\rho(H=0)} \propto \left(\frac{H}{\rho(H=0)}\right)^2 \qquad (S7)$$

At low temperatures, in the R-FL state, $\rho(H=0)$ is estimated by fitting the measured $\rho(H)$ data over a limited magnetic-field range (Fig. 2c), and therefore the resultant value $\rho(H=0)$ may have a considerable error. The magnetic field range for this analysis is constrained by the maximum attainable field and the fact that the upper critical field $H_{c2}$ of the cuprates is very high. In order to improve our estimate of $\rho(H=0)$ in the temperature range in which Kohler's rule is obeyed (Fig. 2f), we simultaneously fit all the $\rho(H)$ data. To this end, we write Eq. (S7) as:

$$\rho(H,T) = \rho(H=0,T) + \frac{\alpha}{\rho(H=0,T)} H^2 \qquad (S8)$$

Where $\alpha$ is a proportionality coefficient that stems from relation (S7). From Eq. (S8), one can see that, if the Kohler scaling is obeyed, the slope of the $H^2$ term is inversely proportional to the (extrapolated) zero-field resistivity. A simultaneous fit of all the data that obey Kohler's rule will therefore increase the precision of the estimation of $\rho(H=0)$. Furthermore, since the extrapolated resistivity exhibits $T^2$ behaviour, we can substitute $\rho(H=0,T) = \rho_{res} + A_2 T^2$ to obtain:

$$\rho(H,T) = \rho_{res} + A_2 T^2 + \frac{\alpha}{\rho_{res} + A_2 T^2} H^2 \qquad (S9)$$

Now, the decrease of the slope of the magnetoresistivity with increasing temperature is clear, although in cases with relatively large residual resistivity, this change might not be very apparent.

For the purpose of the fit, we used the following modification of Eq. S9 (allowing $\rho(H=0,T)$ to be a free fit parameter at each temperature):

$$\rho(H,T) = \rho(H=0,T) + \frac{\beta}{a + T^2} H^2 \qquad (S10)$$

The resultant number of fit parameters equals the number of temperatures at which magnetoresistivity data were collected plus two parameters, $a$ and $\beta$, that are common at all temperatures. The result of the fit is presented in Figs. 2c,e. We note that $a \equiv \frac{\rho_{res}}{A_2}$ and that $a_\perp$, which is defined via $\frac{\delta\rho}{\rho(H=0)} \equiv a_\perp H^2$ (see Fig. S4) in the fit procedure, takes the form:

$$a_\perp = \frac{\beta/\rho(H=0,T)}{a + T^2} \qquad (S11)$$

Since $\rho(H=0,T)$ exhibits a quadratic temperature dependence ($\rho_{res} + A_2 T^2$), Eq. (S11) can be expressed as $a_\perp \propto (d + eT^2)^{-2}$, which is presented in the inset to Fig. 2f.

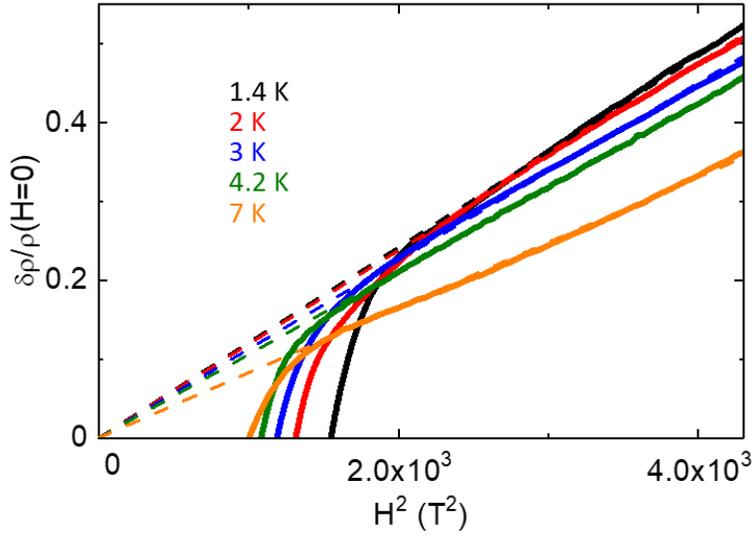

**Figure S4 | Transverse magnetoresistance coefficient $a_\perp$ in Hg1201 (UD71).** Linear functions defined as $F(H,T) = a_\perp H^2$ are plotted alongside $\frac{\delta\rho}{\rho(H=0)}$ vs $H^2$. Excellent agreement is obtained above $H_{c2}$, where the data are fitted. The temperature dependence of $a_\perp$ is presented in the inset to Fig. 2f.

### (S6) Magnetic-field-induced three-dimensional CDW order

Two-dimensional (2D) quasi-static short-range CDW correlations in Hg1201 and Y123 first appear at a temperature ($T_{CDW}$) that is about an order of magnitude higher than the characteristic temperature ($T_{FSR}$) associated with the sign-change of the Hall constant and the peak in the resistivity, and thus with the FS reconstruction (Figs. 1 and 3). As shown most clearly for Y123, in the absence of an applied magnetic field, the CDW correlations increase with decreasing temperature, but then weaken again below $T_c$, due to the competition with superconductivity.[70] On the other hand, CDW correlations are enhanced when superconductivity is suppressed by a magnetic field. Our results are fully consistent with a reconstruction of the pseudogapped FS (*i.e.*, the Fermi arcs) into an electron pocket, caused by bi-directional 2D CDW order in high magnetic fields. Although it has been shown that strong magnetic fields induce additional three-dimensional (3D) CDW order in Y123,[37,38] these charge correlations have not been seen in other cuprates, and it is unlikely that they are linked to the universal FS reconstruction. For example, sensitive ultrasound propagation measurements of Y123 demonstrated that the doping dependence of the 3D CDW order onset temperature does not follow the onset of the FS reconstruction deduced from the sign change of the Hall coefficient,[71] indicating that indeed the 3D correlations are not directly tied to the changes of the FS topology discussed here. Importantly, the wavevector magnitude of the planar CDW order observed in the absence of a magnetic field in structurally simple single-layer Hg1201 is consistent with the reconstruction of Fermi arcs into an electron pocket.[17,18,21] A QO study of Hg1201 unambiguously demonstrated the existence of a single oscillation frequency in the R-FL state, and therefore of a single electron pocket.[21] These results indicate that it is the universal planar 2D CDW correlations that cause the FS reconstruction associated with the transition from the PG-FL to the R-FL state, which requires a complete suppression of superconductivity.

**(S7) Charge localization and the phase diagram**

Across the cuprate phase diagram, the underlying FS contains $1 + p$ states, as seen from photoemission spectroscopy measurements.[11,43,26] In the highly overdoped regime, the underlying FS and actual FS coincide. With decreasing doping, a partial gap develops on the antinodal parts of the FS that is associated with the localization of exactly one hole per $CuO_2$ unit, leading to the formation of Fermi arcs. The resultant change of the effective carrier density from $1 + p$ to $p$ is apparent from dc-resistivity,[11,12,26,40] Hall coefficient,[11,13,40,46] optical conductivity,[15,16] and scanning tunnelling microscopy[72] measurements. Whereas the arc states that remain on the FS preserve their OD-FL character,[11,12,14,15] the antinodal states are associated with (Mott)localization that drives the development of local Cu moments and antiferromagnetic correlations. An unusual aspect of the cuprates is that the evolution from the metal to the insulator is very gradual, with dual *local* (*i.e.*, orbital-selective) Fermi-liquid and pseudogap characteristics. At a qualitative level, this may be understood by considering the distinct possibility of an inherent lack of lattice translational symmetry of the quintessential $CuO_2$ planes.[26,73,74,75] Due to the inhomogeneity, the PG formation is a percolative process that is nearly complete at the characteristic temperature $T^*$ (Fig. S5a); thus, the SM and PG-FL regimes are very similar in character, except that in the PG-FL regime the arcs are fully formed.[26] Analogous to the formation of local Cu moments, the broken-symmetry states observed below $T^*$ ($\boldsymbol{q} = 0$ magnetic order,[76,77] electronic nematicity,[78,79,80] and superconductivity) ought to be seen as emergent phenomena related to the charge localization, which is also a $\boldsymbol{q} = 0$ process.[26]

Regarding superconductivity, it was recently demonstrated that the superfluid density, $\rho_s$, is essentially proportional to *both* the density $n_{\text{eff}}$ of itinerant holes (in electron-[45] and hole-doped cuprates[81]) *and* the density $n_{\text{loc}}$ of localized holes:[26]

$$\rho_s \propto n_{\text{eff}} \cdot n_{\text{loc}}. \tag{S12}$$

Given the constraint

$$1 + p = n_{\text{eff}} + n_{\text{loc}}, \tag{S13}$$

the densities of both electronic subsystems can be directly identified from the normal-state charge transport properties.[11,12,26,42] Figure S5b shows the estimated zero-temperature evolution of $n_{eff}$ determined from both the normal-state resistivity $(n_\rho)$[26] and Hall effect $(n_H)$.[46,42]

The phenomenological observations captured by Eqs. (S12) and (S13) offer a solution to many aspects of the long-standing cuprate high-$T_c$ problem.[11,26,82] For example, Eq. (S12) implies that the localized hole provides the pairing glue, whereas the itinerant (FL) carriers form Cooper pairs and enter the superconducting condensate. Nevertheless, the details related to the microscopic mechanism for superconductivity remain elusive

The presented results provide in this respect very important clues; not only allow a complete characterization of the R-FL state proving the existence of a transition, but also show that this weak CDW order affecting only 0.03 holes per $CuO_2$ unit, is extremely efficient in suppressing superconductivity, presumably by breaking the symmetry of oxygen orbitals.[26,83] Similarly, the low-temperature-tetragonal phase in lanthanum-based cuprates is associated with a small structural change that modifies the relative distances between Cu and different planar O atoms sharply suppressing $T_c$, while barely affecting the normal state transport properties.[84] Finally, it was demonstrated that Zn impurity in the copper-oxygen plane, locally breaks the oxygen symmetry dramatically suppressing $T_c$.[85] Thus, all three experiments indicate that the oxygen-oxygen charge transfer within the unit cell is important, and that a fluctuating oxygen-copper-oxygen polarization could play an essential role in providing the glue.[84] However, more experiments that directly probe the intra-unit-cell physics are required to fully elucidate the superconducting mechanism as well as other aspects related to the localized hole.

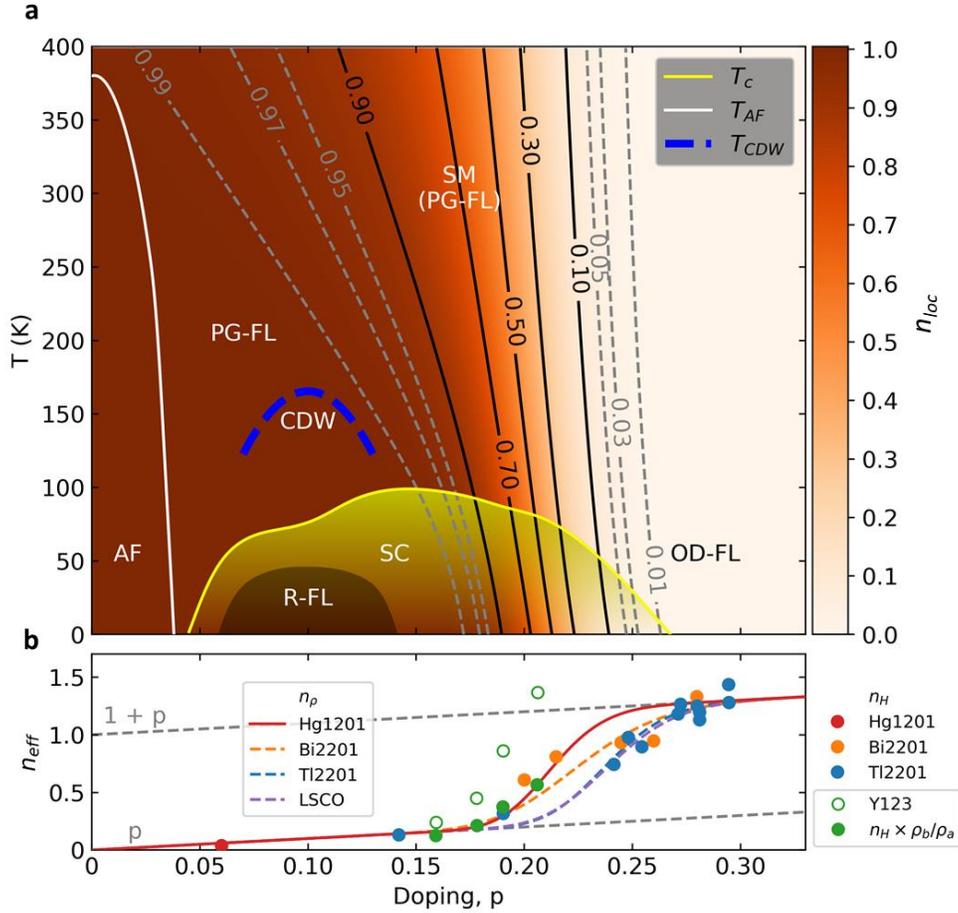

**Figure S5 | Effective charge-carrier density and schematic phase diagram. a,** The colour scale represents the evolution of effective localized carrier density ($n_{loc}$) extracted from resistivity measurements for Hg1201 in the $p \sim 0.06 - 0.18$ range and schematically extended to lower and higher doping based on results for other cuprates.[26] The solid black and dashed grey iso-lines show increments, as indicated. The superconducting (SC), antiferromagnetic (AF), and reconstructed Fermi-liquid (R-FL) phases, and the onset of the short-range charge-density-wave (CDW) correlations are shown. Notably, PG-FL and SM are nearly indistinguishable; in the SM regime, charge localization is not yet fully complete. $T^*$ shown in Fig. 1 (the so-called the PG temperature), above which the temperature dependence of planar resistivity is linear-like, corresponds to ~97% localized carriers, whereas $T^2$ behaviour is observed below $T^{**}$, once charge localization is complete (> 99%). **b**, Zero-temperature effective carrier density $n_{eff}$, extracted from Hall effect ($n_H$) and sheet resistance ($n_\rho$). The data for Tl2201 (blue) and Bi22201 (orange) show a gradual crossover of $n_H$ from $p$ to $1 + p$.[46] Although the data for Y123 (green open circles) appear to indicate a sudden change in carrier density,[86] the same data scaled by the resistivity anisotropy $\rho_b/\rho_a$ (filled green circles, Ref. 46) reveal a more gradual change, consistent with the observations for the other cuprates. The lines indicate $n_\rho$, the effective itinerant normal-state carrier density extracted from resistivity above $T_c$ and extrapolated to $T = 0$ K,[26] taking into account that $m^*/\tau$ is nearly doping and compound independent.[11] Considering the uncertainties in the absolute values of $\rho$ and $R_H$, and the difficulties associated with ascertaining the exact doping level for each particular sample, the agreement between $n_\rho$ and $n_H$ among different compounds is remarkable.

## (S8) Resistivity and Hall coefficient in very underdoped Hg1201

Our work confirms the notion[11,12] that the normal state from which superconductivity emerges on cooling is characterized by the dual properties of localized charge and itinerant Fermi-liquid quasiparticles that persist on arcs (PG-FL). It is an intriguing question if, in the absence of superconductivity and significant CDW correlations, the PG-FL persists to zero temperature. This has proven to be a challenging experimental task. On the one hand, the quasi-static CDW correlations span a wide doping range, cover a substantial part of the PG-FL regime, and drive the emergence of a new, field-induced state with a reconstructed FS. On the other hand, around optimal doping ($p \approx 0.16$), the SC state is particularly robust against applied magnetic fields, with $H_{c2}$ greater than 100 T in Hg1201[25] and 200 T in Y123.[87] Although efforts have been made to study the electronic transport down to very low temperatures at the edge of the SC dome, below the doping range where the CDW order is observed,[88,89] in most cuprates the intrinsic transport behaviour is masked by non-universal (approximately logarithmic) resistivity upturns.[39] These upturns were shown to be compound specific,[39] promoted by disorder,[90] and are probably related to the percolative appearance of the insulating phase.[14,39]

However, in a very underdoped sample of Hg1201 ($T_c$ = 45 K, $p \approx 0.06$), no such upturn was observed.[25] We now provide our own analysis of these data. Notably, no CDW order has been observed at this doping level in Hg1201.[18] Even without a detailed analysis, or considering sample-related issues (*e.g.*, $T_c$ extracted from resistivity is much higher than the value determined from magnetic susceptibility), the data in Ref. 25 indicate that the PG-FL extends down to the lowest measured temperature of ~10 K, and to fields higher than 64 T. Namely, the quadratic temperature dependence of the resistivity appears to extend all the way from $T^{**} \approx 230$ K to very low temperatures (Fig. S6a). This value of $T^{**} \approx 230$ K, the temperature above which the planar resistivity starts to deviate from $T^2$, is consistent with previous results.[12,11] At low temperatures and high fields, a rather small deviation from the quadratic high-temperature zero-field fit is seen (Fig. S6a) due to the fact that the magnetoresistance increases with increasing field and decreasing temperature. Unfortunately, from the data in Ref. 25, it is impossible to properly subtract this magnetoresistance due to the apparent presence of filamentary superconductivity.

The Hall resistance data in Ref. 25 provide further insight into the underlying non-SC ground state. In the normal state, $R_{xy}$ exhibits a linear dependence on magnetic field, with the zero intercept at $H$ = 0 T (Fig. S6b). This implies that $R_H$, and thus $n_H$, is independent of magnetic field. Furthermore, as shown in Fig. S6c, $R_H$ remains approximately constant in the temperature range in which the magnetic field of 64 T suppresses superconductivity, while simultaneously the extrapolated intercept of $R_{xy}$ is negligible. At lower temperatures, where the magnetic field is insufficient to fully suppress superconductivity (attributed here to the presence of filamentary SC), the $R_H$ isotherms asymptotically approach the $H$ = 0 value of $R_H \approx 27.7$ mm$^3$/C. This indicates that the carrier density is essentially temperature independent down to the lowest temperatures (at 4 K, the SC contribution is too strong to be considerably suppressed by the applied field).

Consequently, the data of Ref. 25 strongly suggest that, in the absence of CDW order and superconductivity, the PG-FL with its Fermi-arc topology extends to zero-temperature.

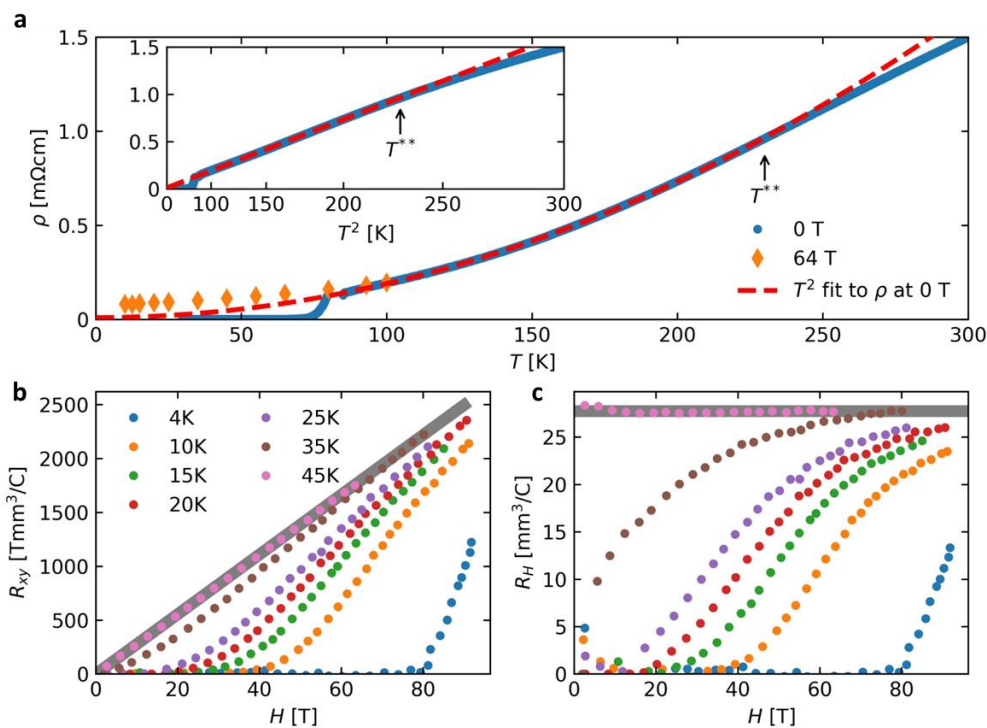

**Figure S6 | Resistivity and Hall coefficient in strongly underdoped Hg1201 ($T_c$ = 45 K, $p \approx 0.06$) from Ref. 25. a,** Temperature dependence of the planar resistivity in the absence of a magnetic field (blue dots), and in a 64 T $c$-axis field (orange diamonds). The data are reproduced from Fig. 2b in Ref. 25 and, for consistency, rescaled to match the previously-reported resistivity.[11] The red dashed line is a quadratic fit to $\rho(H=0)$ in the 100 - 200 K range; the fit describes the data up to $T^{**} \approx 230$ K. The high-field data also follow well $T^2$ dependence, with a small (expected) upward deviation due to magnetoresistivity. The inset presents the zero-field data on a quadratic scale. **b**, Hall resistance as a function of field at indicated temperatures. The grey line corresponds to a linear fit to the 45 K data. Remarkably, all other isotherms, except the one at 4 K, asymptotically approach the 45 K linear dependence at high field. At 4 K, the maximum applied field is still insufficient to fully suppress the SC. **c,** Hall coefficient as a function of magnetic field, calculated from the data in **b**. The grey line corresponds to the slope of the fit in **b** and equals $R_H$ = 27.7 ± 0.7 mm$^3$/C. This value corresponds to $p \approx 0.04$, which is slightly below the value $p \approx 0.06$ estimated based on an empirical relation between $T_c$ and thermoelectric power.[54] At temperatures below 35 K, where superconductivity is not completely suppressed by the magnetic field, $R_H$ asymptotically approaches the value measured at higher temperatures. Together with previously reported Hall data at temperatures closer to $T^*$,[11] this indicates that $R_H$ is essentially temperature independent within the PG-FL regime down to the lowest measured temperatures.

### (S9) Arc-to-hole-pocket transformation

A recent combined photoemission and QO study focused on $Ba_2Ca_4Cu_5O_{10}(F,O)_2$, a five-layer cuprate characterized by strong antiferromagnetism.[44] The photoemission experiments revealed that, in addition to the Fermi arcs, the FS features two hole pockets. The hole pockets were found to correspond to the inner $CuO_2$ planes and to a relatively low carrier concentration (below 5%) due to screening from the outermost layers, whereas the arcs were found to be associated with the relatively highly doping (above 5%) of the outer layers. The simultaneous observation of pockets and arcs has two important implications. First, the ground-state FS near the low-doping edge of the SC state consists of Fermi arcs, consistent with our analysis in Section (S8). Second, the AF correlations at very low doping (Fig. 1a) fold the arcs into hole pockets[44] in a manner similar to the FS reconstruction into an electron pocket due to CDW correlations at higher doping in the presence of external magnetic field.[17,18,21,22]

Finally, we make another observation. The transport measurements of $Ba_2Ca_4Cu_5O_{10}(F,O)_2$ revealed only the two QO frequencies associated with the two hole pockets; no QOs associated with the arcs were detected.[44] Although the photoemission data indicate that the scattering on arcs is somewhat higher than on the pockets, this observation indicates that arcs cannot be observed in QO measurements. This in turn provides a plausible explanation for the lack of evidence for QOs in overdoped Tl2201 below $p \sim 0.28$,[91] since we would expect the onset of arc formation due to the opening of the PG at about this doping level. This is seen from Fig. S5b, *i.e.,* from the decrease of the carrier density indicated by both the resistivity and the Hall effect.[11,12,26,46]

## Supplementary References